\documentclass[10pt,twocolumn,floatfix,prb,aps,showpacs]{revtex4-1}
\usepackage{graphicx}
\usepackage{amssymb}
\usepackage{amsmath}
\usepackage{color}
\usepackage{psfrag}
\usepackage{epsfig}
\usepackage{epstopdf}
\usepackage{bbm}
\usepackage{bm}
\usepackage{hyperref}
\hypersetup{breaklinks}

\newcommand{\beq}{\begin{eqnarray}}
\newcommand{\eeq}{\end{eqnarray}}
\newcommand{\eq}[1]{Eq.~(\ref{#1})}
\newcommand{\fig}[1]{Fig.~\ref{#1}}
\newcommand{\rb}[1]{\left( #1 \right)}

\newcommand{\citer}[1]{{Ref.~\onlinecite{#1}}}
\newcommand{\secref}[1]{{Sec.~\ref{#1}}}

\begin{document}
\title{ Coherent feedback control in quantum transport}
\author{Clive Emary}
\affiliation{
  Department of Physics and Mathematics,
  University of Hull,
  HU6 7RX,
  United Kingdom
}
\author{John Gough}
\affiliation{
  Department of Mathematics and Physics,
  Aberystwyth University, 
  SY23 3BZ,
  United Kingdom
}
\date{\today}
\begin{abstract}
  We discuss control of the quantum-transport properties of a mesoscopic device by connecting it in a coherent feedback loop with a quantum-mechanical controller.
  We work in a scattering approach and derive results for the combined scattering matrix of the device-controller system and determine the conditions under which the controller can exert ideal control on the output characteristics.
  As concrete example we consider the use of feedback to optimise the conductance of a chaotic quantum dot and investigate effects of controller dimension and  decoherence.
  In both respects we find that the performance of the feedback geometry is well in excess of that offered by a simple series configuration.
\end{abstract}
\maketitle

\section{Introduction}

Feedback is one of the fundamental techniques of classical control theory \cite{Bechhoefer2005} and its translation into the quantum realm \cite{Wiseman2009, Gough2012} seems certain to play an equally important role in the rapidly-developing field of quantum technology.
This work concerns itself with the application of feedback control to quantum transport, where a number of interesting effects have already been predicted, such as the freezing of current fluctuations \cite{Brandes2010}, stabilisation of quantum states \cite{Kiesslich2011,Poeltl2011,Kiesslich2012}, realisation of a mesoscopic Maxwell's daemon \cite{Schaller2011, Esposito2012,Strasberg2013} and delay effects \cite{Emary2013b}.
In these hitherto-proposed schemes, the feedback loops employed were examples of measurement-based quantum control \cite{Wiseman1994,Wiseman2009}, in which the full counting statistics of electron transport \cite{Levitov1996,Bagrets2003,Gustavsson2006,Fujisawa2006} were monitored and control operations applied to the system in response to individual electron tunnelling events.
The feedback loop in such cases is classical, as is the information to flow between system and controller.

In contrast, we are here interested in the application of {\em coherent feedback control} to quantum transport.
In coherent control, the system (or, to borrow the engineering term, the {\em plant}), the controller and their interconnections are all quantum-mechanical and phase coherent.  The system-controller complex therefore evolves under a joint unitary dynamic and the information flow between plant and controller is of quantum, rather than classical, information \cite{Lloyd2000}.
The main advantages of coherent feedback control over its measurement-based cousin are held to be\cite{Zhang2011}:  reduced noise, since the additional disturbance produced by the quantum-mechanical measurement process is absent; and speed, since the coherent controller is likely to operate on the same time scales as the plant (in contrast, a classical controller will be limited to speeds associated with traditional electronics).

Various forms of coherent control have been discussed in the literature, e.g. Refs.~\onlinecite{Lloyd2000,Mabuchi2008,Zhang2011}, but the type we will focus on here is the {\em quantum feedback network} developed by one of the authors with James \cite{Gough2008,Gough2009,Gough2009a,James2008,Nurdin2009,Zhang2011} (see \citer{Zhang2012} for a recent review). 
The proposal of such networks can be traced to the cascading of open systems due to Carmichael and Gardiner \cite{Carmichael1993a,Gardiner1993}, feedback connections for linear quantum systems \cite{Yanagisawa2003}, as well as the all-optical measurement-based feedback schemes of Wiseman and Milburn \citer{Wiseman1994a}, which can also be placed in this setting \cite{Gough2009a}. A number of experiments have been performed in this paradigm, including disturbance rejection \cite{Mabuchi2008} and the control of optical squeezing \cite{Iida2012,Crisafulli2013}.  Further proposals include automatic quantum error correction \cite{Kerckhoff2010,Kerckhoff2011}, suppression of switching in bistable optical systems \cite{Mabuchi2011}, cavity cooling \cite{Hamerly2012,Hamerly2013}, and the generation of entangled photons \cite{Hein2014}.

\begin{figure}[t]
  \psfrag{S}{\scalebox{3}{$S$}}
  \psfrag{S0}{\scalebox{3}{$S_0$}}
  \psfrag{K}{\scalebox{2.7}{$K$}}
  \psfrag{K0}{\scalebox{2.3}{$K_0$}}
  \psfrag{K1}{\scalebox{1.7}{$K_1$}}
  \psfrag{K2}{\scalebox{1.7}{$K_2$}}
  \psfrag{A}{\scalebox{1.5}{A}}
  \psfrag{B}{\scalebox{1.5}{B}}
  \psfrag{C}{\scalebox{1.5}{C}}
  \psfrag{D}{\scalebox{1.5}{D}}
  \psfrag{N}{\scalebox{1}{$N$}}
  \psfrag{M}{\scalebox{1}{$M$}}
  \psfrag{N'}{\scalebox{1}{$N'$}}
  \psfrag{(a)}{\scalebox{1}{\textbf{(a)}}}
  \psfrag{(b)}{\scalebox{1}{\textbf{(b)}}}
  \psfrag{(c)}{\scalebox{1}{\textbf{(c)}}}
  \begin{center}
     \includegraphics[width=\columnwidth,clip=true]{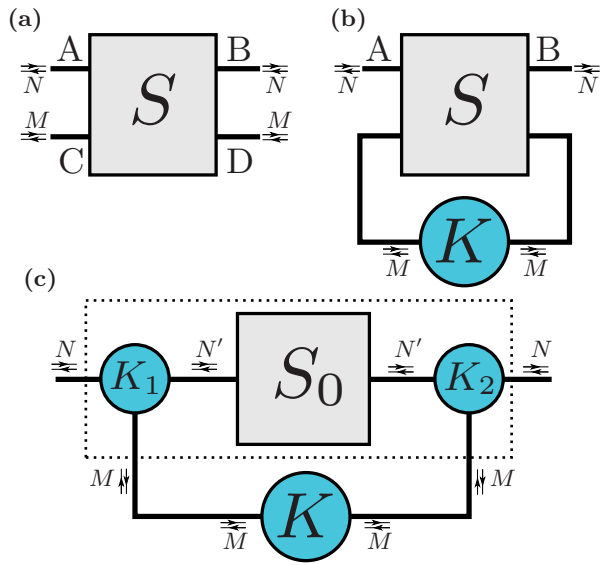}
  \end{center}
  \caption{(color online)
    Schematics of quantum feedback network consisting of a mesoscopic device, $S$, and controller, $K$.
    \textbf{(a)}
    The isolated device with four leads labelled A through D.  Leads A and B possess $N$ (bidirectional) channels; leads C and D possess $M$.
    \textbf{(b)}
    The feedback loop is realised by connecting the leads C and D together  via the controller.
    After feedback, the device becomes a scatterer between the $N$ channels of lead $A$ and the $N$ channels of lead B.
    \textbf{(c)}
    A feedback network where the original device is a two-terminal device, $S_0$.  The addition of scatterers $K_1$ and $K_2$ converts $S_0$ into a four-terminal device (enclosed in the dashed box here), such that this network maps on to that in part (b).
    \label{FIG:SFB}
  }
\end{figure}

Whilst these developments have taken place largely in the context of quantum optics, our aim here is to study  coherent feedback control in quantum transport.  In particular, we are interested in how a quantum feedback network can be used to modify the conduction properties of a mesoscopic device.
To be specific, our focus will be on four-terminal devices, \fig{FIG:SFB}a, which we embed in a feedback network by connecting two of the four leads in a loop via some external control circuit or device, \fig{FIG:SFB}b.  
We will assume that the  motion of electrons through plant and controller is phase coherent and that electron-electron interactions can be neglected.  In this limit, transport can be described by Landauer-B\"uttiker theory \cite{Blanter2000}, where both the plant and controller are described by scattering matrices.

Analysis of the feedback loop amounts to finding the composite scattering matrix of the system-controller complex and relating this to the conduction properties.
From this, the main formal result is that when the number of controller channels, $M$, equals the number of plant channels that remain after feedback, $N$, then free choice of the control scattering matrix allows us to set the scattering matrix of combined system as desired.  We refer to this situation as ``ideal control'', and since the scattering matrix can be set at will, so can the conduction properties (within natural limits set by the dimensionality of the scatterer).

We then explore the issue of what happens away from this ideal limit.  The first question we address is to what extent can the conductance be controlled when the size of the controller is lower than required for ideal control, i.e. when $M < N$.
To answer this we consider the concrete example of a chaotic quantum dot, the scattering through which we describe with random matrix theory \cite{Beenakker1997}.  We assume a unconstrained controller and choose its parameters so as to optimise the conductance through the dot as a function of $0 \le M \le N$.
We compare these results with those obtained from a second control geometry in which the quantum dot is connected to the controller in series, \fig{FIG:series}.  
We find that, for all $0<M<N$, the feedback geometry significantly outperforms the series for conduction maximisation.

Secondly, we consider the effects of dephasing on both feedback and series geometries and show that conduction maximisation in the feedback case is far more robust to dephasing than the series case.  
Indeed, the feedback geometry can provide some degree of conductance increase even in the presence of total dephasing.  The series geometry cannot.

This paper proceeds as follows. In \secref{SEC:scat}, we discuss the scattering problem and derive an expression for the scattering matrix of combined system-controller network in both feedback and series geometries.
\secref{SEC:ideal} introduces the notion of ideal control and examines the conditions under which it can pertain. 
Secs.~\ref{SEC:QD} and \ref{SEC:deph} discuss numerical results for conductance optimisation for the quantum dot and focus on the effects of controller dimension and dephasing, respectively.  Finally, 
\secref{SEC:disc} contains some concluding remarks and perspectives.

\section{Quantum feedback network \label{SEC:scat}}

The plant here is a four-terminal mesoscopic conductor with leads labelled A and C on the left and B and D on the right (\fig{FIG:SFB}a).   Leads A and B each support $N$ conduction channels; leads C and D each support $M$
\footnote{
  We describe two groupings of channels on each side (e.g. A and C on the left) as inhabiting physically distinct leads, but this need not necessarily be the case.  The key requirement is that the two groups of channels be  independently accessible.
}.
Should the plant of interest actually be a two-terminal conductor, use can be made of the geometry shown in \fig{FIG:SFB}c.  Here two three-terminal scatterers,
presumably very simple, are added before and after the original two-terminal plant.
The composite of these three elements is then a four-terminal device, as assumed by the following formalism
\footnote{
  The layout \fig{FIG:SFB}c is reminiscent of the canonical feedback loop of classical control theory.  The important thing to realise is that here the interconnects represent leads that support transport in both directions, rather than transfer in just a single direction.  Thus, there is no issue with the cloning of quantum information here, as there is in the straightforward quantum generalisation of the classical transfer model.
}.

Let $b^\mathrm{in}_{\mathrm{X}, n}(E)$ be the annihilation operator for an incoming electron of energy $E$ in channel $n$ of lead $\mathrm{X=A,B,C,D}$, and let $b^\mathrm{out}_{\mathrm{X},n}(E)$ be the corresponding operator for an outgoing electron.  In Landauer-B\"uttiker theory \cite{Blanter2000}, the device is treated as a phase-coherent scatterer of electrons with incoming and outgoing states related by
\beq
  {b}^\mathrm{out}(E) = S(E) \;  b^\mathrm{in}(E)
  ,
  \label{eq:io}
\eeq
where $b^\mathrm{in}(E)$ and $b^\mathrm{out}(E)$ are vectors containing the appropriate annihilation operators of all leads, and $S(E)$ is the scattering matrix of the device at energy $E$.
Note that the scattering matrix $S$ must be unitary.
In the current work we will consider linear transport only and, in this case, the only relevant energy is the Fermi energy of the leads A and B. All quantities, in particular the scattering matrix $S$, will be evaluated at this Fermi energy, and we suppress the energy index from now on.

We write \eq{eq:io} as
\beq 
  \rb{
    \begin{array}{c}
      b^\mathrm{out}_{\mathrm{A}} \\
      b^\mathrm{out}_{\mathrm{B}} \\
      b^\mathrm{out}_{\mathrm{C}} \\
      b^\mathrm{out}_{\mathrm{D}} 
    \end{array}
  }
  =
  \rb{
    \begin{array}{cccc}
      S_{\mathrm{AA}} & S_{\mathrm{AB}} & S_{\mathrm{AC}} & S_{\mathrm{AD}} \\
      S_{\mathrm{BA}} & S_{\mathrm{BB}} & S_{\mathrm{BC}} & S_{\mathrm{BD}} \\
      S_{\mathrm{CA}} & S_{\mathrm{CB}} & S_{\mathrm{CC}} & S_{\mathrm{CD}} \\
      S_{\mathrm{DA}} & S_{\mathrm{DB}} & S_{\mathrm{DC}} & S_{\mathrm{DD}} \\
    \end{array}
  }
   \rb{
    \begin{array}{c}
      b^\mathrm{in}_{\mathrm{A}} \\
      b^\mathrm{in}_{\mathrm{B}} \\
      b^\mathrm{in}_{\mathrm{C}} \\
      b^\mathrm{in}_{\mathrm{D}} 
    \end{array}
  }
  ,
  \; \; \;  \; 
  \label{eq:io_matrix}
\eeq
where the component $S_{\mathrm{XY}}$ is the matrix relating the input to lead Y (i.e., $b^\mathrm{in}_\mathrm{Y}$)
with the output from lead $X$  (i.e., $b^\mathrm{out}_\mathrm{X}$).

\begin{figure}[tbp]

	\flushleft{\hspace{5mm}\textbf{(a)}}
	
	\includegraphics[width=\columnwidth]{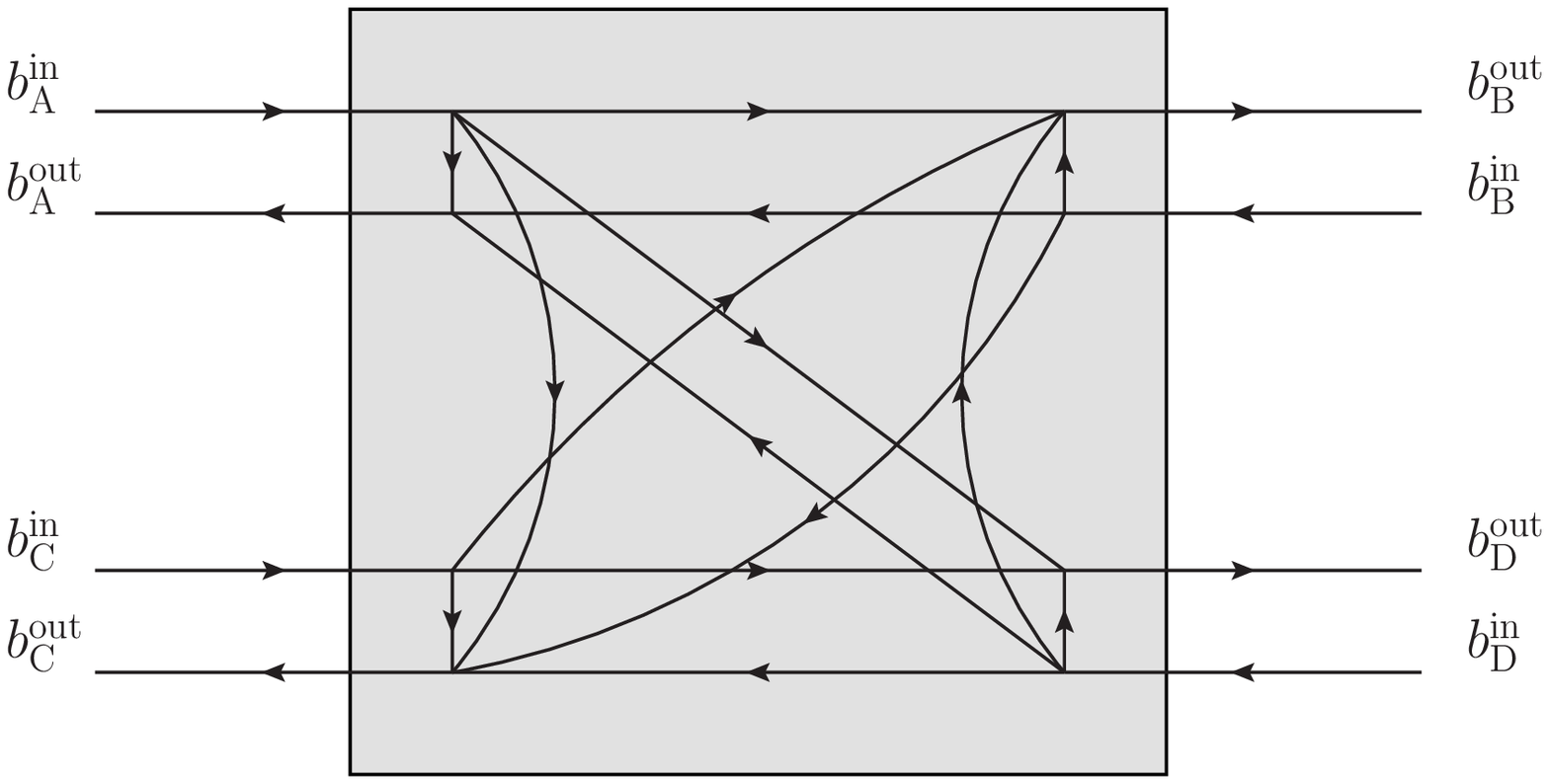}
	
	\flushleft{\hspace{5mm}\textbf{(b)}}
		
	\includegraphics[width=\columnwidth]{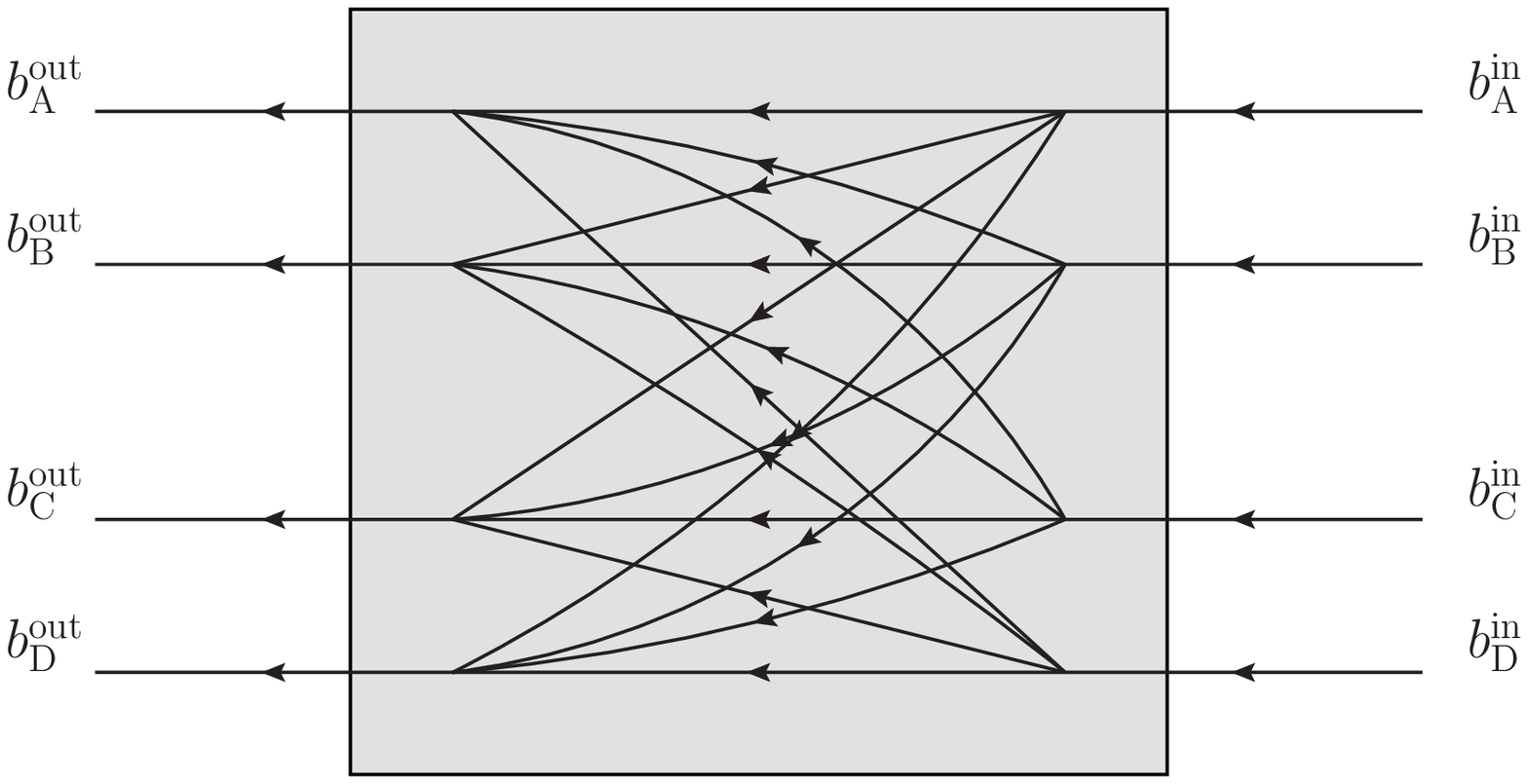}
	\caption{
	  Two equivalent representations of an input-output device describing the scattering of fields 
	  $b_\mathrm{X}^\mathrm{in}$ into 
	  $b_\mathrm{X}^\mathrm{out}$ in leads $\mathrm{X=A,B,C,D}$.
	  These fields are multidimensional with $N$ being the multiplicity of modes  
	  $b^\mathrm{in}_{\mathrm{A}}$,
	  $b^\mathrm{out}_{\mathrm{A}}$, 
	  $b^\mathrm{in}_{\mathrm{B}}$, and
	  $b^\mathrm{out}_{\mathrm{B}}$,  
	  and with $M$ being the multiplicity of modes 
	  $b^\mathrm{in}_{\mathrm{C}}$,
	  $b^\mathrm{out}_{\mathrm{C}}$, 
	  $b^\mathrm{in}_{\mathrm{D}}$, and
	  $b^\mathrm{out}_{\mathrm{D}}$.
	  In \textbf{(a)},
	  as in \fig{FIG:SFB}a, fields within a given lead are grouped together, with leads A and C on the left and B and D on the right.
	  In this representation, the device is seen to be a four-lead device, where each lead is bidirectional.
	  \textbf{(b)} shows a representation that mirrors the action of the scattering matrix in which all input fields are drawn to the right, and all output fields leave on the left.
	  \label{FIG:binary_io}
	}
\end{figure}

In \fig{FIG:binary_io} we show two representations of the scattering by this device.  In the first representation the modes are organised in terms of the physical leads (e.~g. $b^\mathrm{in}_{\mathrm{A}}$ and $b^\mathrm{out}_{\mathrm{A}}$ are grouped together); the second representation reflects the structure of the scattering matrix.

\subsection{Feedback \label{SUBSEC:FB}}

We introduce feedback by connecting the channels in lead C to those in lead D  via the controller.  This latter we describe with the $2M \times 2M$ dimensional scattering matrix, $K$.
To facilitate our description, we partition the scattering matrix in terms of those channels that will form the feedback loop (those in leads $C$ and $D$) and those  that will persist ($A$ and $B$).  We therefore write
\beq
  S = 
  \rb{
    \begin{array}{cc}
      S_\mathrm{I} & S_\mathrm{II} \\
      S_\mathrm{III} & S_\mathrm{IV}
    \end{array}
  }
  \label{EQ:Sfull}
  ,
\eeq
with blocks 
\beq
  S_\mathrm{I} = 
  \rb{
    \begin{array}{cc}
      S_{\mathrm{AA}} & S_{\mathrm{AB}} \\
      S_{\mathrm{BA}} & S_{\mathrm{BB}}
    \end{array}
  }
  ;\quad
  S_\mathrm{II} = 
  \rb{
    \begin{array}{cc}
      S_{\mathrm{AC}} & S_{\mathrm{AD}} \\
      S_{\mathrm{BC}} & S_{\mathrm{BD}}
    \end{array}
  }
  ;
  \nonumber\\
  S_\mathrm{III} = 
  \rb{
    \begin{array}{cc}
      S_{\mathrm{CA}} & S_{\mathrm{CB}} \\
      S_{\mathrm{DA}} & S_{\mathrm{DB}}
    \end{array}
  }
  ;\quad
  S_\mathrm{IV} = 
  \rb{
    \begin{array}{cc}
      S_{\mathrm{CC}} & S_{\mathrm{CD}} \\
      S_{\mathrm{DC}} & S_{\mathrm{DD}}
    \end{array}
  }
  .
\eeq
The scattering matrix for the complete feedback network, $S_\mathrm{fb}$,
can then be derived by considering all scattering processes between leads A and B.
Firstly, there is direct scattering, which is described by scattering block $S_\mathrm{I}$.  Electrons can also be scattered into the feedback loop, traverse it once, and then reemerge into the ``AB system''.  This is described by the sequence of matrices  
$S_\mathrm{II} K S_\mathrm{III} $.
Further processes are then possible in which the electron makes $n$ traversals of the feedback loop, to give the scattering term 
$
  S_\mathrm{II}
 \rb{ K S_\mathrm{IV}}^{n}
  K S_\mathrm{III}
  ;~ n=1,2,\ldots
$.  Summing the totality of all possibilities, the total scattering matrix of the system with feedback reads:
\beq
  S_\mathrm{fb}
  =
  S_\mathrm{I}
  +
  S_\mathrm{II}
  \frac{1}{\mathbbm{1}- K S_\mathrm{IV}}
  K S_\mathrm{III}
  \label{EQ:SFB1}
  ,
\eeq
with $\mathbbm{1}$ the unit matrix (here of dimensions $2M\times 2M$).
This form relies on the existence of inverse of $\mathbbm{1}- K S_\mathrm{IV}$ and physically, this corresponds to the condition that all $M$ channels connect through the controller.
Similar results have been derived previously, see, e.~g., \citer{Gough2008}.

\subsection{Series \label{SUBSEC:series}}

To gain an appreciation of the utility of the feedback geometry, we will compare it with a further plant-controller network, namely the {\em bidirectional series} connection: this is the generalization
of the series product for unidirectional fields introduced in \citer{Gough2009a}.

\begin{figure}[tbp]
  \begin{center}
    \includegraphics[width=\columnwidth]{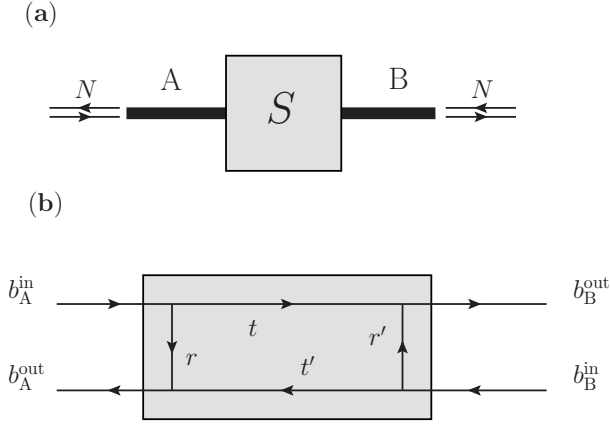}
  \end{center}
    \caption{
	\textbf{(a)} A two-lead device with leads A and B each of which supports $N$ bidirectional channels. Scattering is described by unitary matrix  $S$
	\textbf{(b)} Equivalently, the system can be described as a four-port unidirectional device, where the fields $b^\mathrm{in}_\mathrm{A} ,b^\mathrm{in}_\mathrm{B},b^\mathrm{out}_\mathrm{A} ,b^\mathrm{out}_\mathrm{B}$ each have multiplicity $N$.
    }
    \label{FIG:2port}
\end{figure}

Our first order of business is describe how we model two-port bidirectional systems. These arise as 
unidirectional four port systems, see \fig{FIG:2port} where the inputs $b^\mathrm{in}_\mathrm{A} ,b^\mathrm{in}_\mathrm{B}$
and outputs $b^\mathrm{out}_\mathrm{A} ,b^\mathrm{out}_\mathrm{B}$ each have multiplicity $N$.  The scattering matrix may then be written in block form as
\beq
  S=
  \rb{
    \begin{array}{cc}
      r & t' \\
      t & r'
    \end{array}
  }
  ,
	\label{eq:S-matrix}
\eeq
that is $r=S_\mathrm{AA}$ is the $N\times N$ complex matrix describing the reflection coefficients
of input $b^\mathrm{in}_\mathrm{A}$ in $b^\mathrm{out}_\mathrm{A}$, $t=S_\mathrm{BA}$ describes the transmission coefficients
of $b^\mathrm{in}_\mathrm{A}$ into $b^\mathrm{out}_\mathrm{B}$, etc.

\begin{figure}[tb]
  \begin{center}
    \includegraphics[width=\columnwidth]{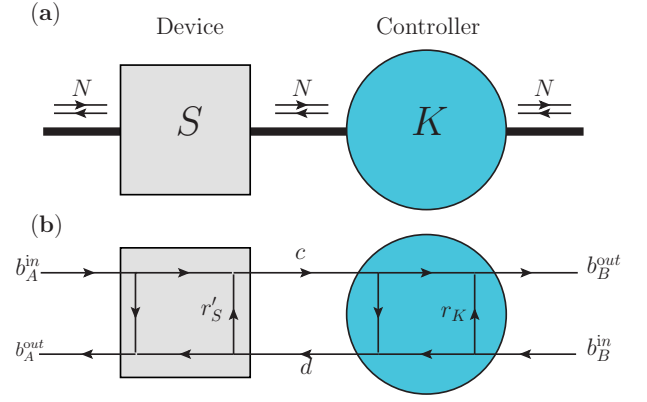}
  \end{center}
  \caption{ 
    (color online)
    \textbf{(a)} A mesoscopic device, $S$, and controller, $K$, connected in the {\em bidirectional series configuration}, which we denote by $S \Diamond K$.  All leads support $N$-channels in each direction.
    \textbf{(b)} The bidirectional series connection of devices $S$ and $K$ may be expressed in terms of unidirectional models where we see explicitly the presence of a single algebraic feedback loop where output $c$ from the plant $S$ is fed into $K$ and contributes to output $d$ with gain $r_K$,
    while $d$ enters $S$ and contributes to output $c$ with gain $r'_S$.
    \label{FIG:series}
  }
\end{figure}

The bidirectional series construction between a plant with scattering matrix $S$ and a controller with scattering matrix $K$ is shown in \fig{FIG:series}.
Here both $S$ and $K$ are $2N\times 2N$ unitary matrices which act between input and output fields as
\begin{eqnarray}
\rb{
    \begin{array}{c}
      b^\mathrm{out}_\mathrm{A} \\
       c
    \end{array}
  }
  =
  \rb{
    \begin{array}{cc}
      r_S  & t^\prime_S \\
      t_S  & r^\prime_S
    \end{array}
  }
  \rb{
    \begin{array}{c}
      b^\mathrm{in}_\mathrm{A} \\
       d
    \end{array}
  } ,
  \nonumber \\
  \rb{
    \begin{array}{c}
      d \\
      b^\mathrm{out}_\mathrm{B} 
    \end{array}
  } 
  =
  \rb{
    \begin{array}{cc}
      r_K  & t^\prime_K \\
      t_K  & r^\prime_K
    \end{array}
  }
  \rb{
    \begin{array}{c}
      c \\
      b^\mathrm{in}_\mathrm{B} 
    \end{array}
  }  
  .
\end{eqnarray}
From this we see that
\begin{eqnarray}
\rb{
    \begin{array}{cc}
      \mathbbm{1}  & -r^\prime_S \\
      -r_K  & \mathbbm{1}
    \end{array}
  }
\rb{
    \begin{array}{c}
      c \\
     d 
    \end{array}
  } 
  =
    \rb{
    \begin{array}{c}
     t_S \, b^\mathrm{in}_\mathrm{A} \\
     t^\prime_K \, b^\mathrm{in}_\mathrm{B} 
    \end{array}
  }  
  ,
\end{eqnarray}
The network will be well-posed if we can invert the matrix to solve for $c$ and $d$. Assuming that this is indeed the case, then we can use the block matrix inversion (Banachiewicz) formula
\begin{eqnarray}
\rb{
    \begin{array}{cc}
      \mathbbm{1}  & -r^\prime_S \\
      -r_K  & \mathbbm{1}
    \end{array}
  }^{-1}
  =
    \rb{
    \begin{array}{cc}
     \Delta_{SK} & \Delta_{SK} \, r^\prime_S \\
     \Delta_{KS} r_K & \Delta_{KS} 
    \end{array}
  }  
  , \; \; \; 
\end{eqnarray}
where
\beq
  \Delta_{SK} = (\mathbbm{1}-r^\prime_{S} r_K )^{-1} , \nonumber \\
  \Delta_{KS} = (\mathbbm{1}- r_K r^\prime_S )^{-1} 
  .
\eeq
This allows us then to write
\beq
 \rb{
    \begin{array}{c}
      b^\mathrm{out}_\mathrm{A} \\
      b^\mathrm{out}_\mathrm{B}
    \end{array}
  }
=
  \rb{
    \begin{array}{cc}
      r & t' \\
      t & r'
    \end{array}
  }
   \rb{
    \begin{array}{c}
      b^\mathrm{in}_\mathrm{A} \\
      b^\mathrm{in}_\mathrm{B}
    \end{array}
  } ,
\eeq
with the blocks
\beq
 r   & = & r_S +t^\prime_S r_K \Delta_{SK} t_S
    \nonumber  \\
 t^\prime   & = & t^\prime_S \Delta_{KS} 
    \nonumber  \\
  t   & = &
  t_K \Delta_{SK}  t_S
  \nonumber
  \\
  r^\prime 
  & = &
  r^\prime_K + t_K r^\prime_S \Delta_{KS} 
  \label{EQ:seriestr}
  .
\eeq
This then is the combined scattering matrix for the bidirectional series product $S \Diamond K$ as defined in \fig{FIG:series}.
The joint scattering matrix, $S_{S\Diamond K}$, obtained this way agrees with standard calculation for two scatterers $S$ and $K$ connected in series \cite{Datta1997}. We remark that the formula should generalize to the 
situation where both the plant and controller are Markovian quantum systems which involve an internal dynamical $H$, coupling/collapse operators $L$, in addition to just the scattering matrix $S$.

To make a direct comparison between series and feedback cases, we construct the control matrix $K$ here to have $N-M$ trivially-transmitting channels and $M$ channels that are actually subject to a control scattering matrix.

\subsection{Conductance}

In the limit of low temperature and small bias about a Fermi energy $E_F$, the conductance of a two-terminal sample with a scattering matrix as in \eq{eq:S-matrix} is given by \cite{Blanter2000}
\beq
  G 
  &=& G_0\mathrm{Tr}[t^\dag t]
  ,
\eeq
where $G_0 =  {\textstyle \frac{2e^2}{h}}$ is the conductance quantum (all channels assumed spin-degenerate) and where the transmission block $t$ is evaluated at the Fermi energy $t = t(E_F)$.
With $T_{n}$ the transmission probabilities given by the eigenvalues of matrix $t^\dag t$, the conductance can be written
\beq
  G = G_0 \sum_n T_{n}
  \label{EQ:cond}
  .
\eeq

\section{Ideal control \label{SEC:ideal}}

When the control scattering matrix has the same dimension as the output matrix, i.e. when $N=M$, the matrices $S_\mathrm{II}$ and $S_\mathrm{III}$ are square. Assuming that the determinants of these two matrices are non-zero (see below) these matrices are {\em invertible} and it becomes possible to rearrange \eq{EQ:SFB1} 
for the control matrix as
\beq
  K = 
  \frac{1}{S_\mathrm{IV} + S_\mathrm{III} \rb{S_\mathrm{fb}-S_\mathrm{I}}^{-1}S_\mathrm{II}}
  \label{EQ:Kideal}
  .
\eeq
Thus, given an arbitrary plant matrix $S$, we can obtain any given target $S_\mathrm{fb}$ by choosing the control operator as in \eq{EQ:Kideal}.  And if $S_\mathrm{fb}$ can be chosen arbitrarily, so can the transmission eigenvalues $T_n$ and all desired conductance properties.
The inversion of $S_\mathrm{fb}$ to obtain \eq{EQ:Kideal} requires that the inverses $S_\mathrm{II}^{-1}$, $S_\mathrm{III}^{-1}$ and $\rb{S_\mathrm{fb}-S_\mathrm{I}}^{-1}$ exist.  Physically, the absence of these inverses corresponds to the case when one or more of the channels in leads A or B are completely decoupled from leads C and D. In this case, it is clear that  these modes can not be affected by the feedback loop and thus ideal control is not possible.

The possibility of ideal control also exists for the series case.  Provided that the inverses $t_S^{-1}$ and ${t_S'}^{-1}$ exist, equation set~(\ref{EQ:seriestr}) can be inverted to obtain the ideal control matrix $K$.  As above, this requires that the number of channels in control and output spaces be equal, $M=N$.

\section{Conductance optimisation of a chaotic quantum dot \label{SEC:QD}}

When the dimension of the controller equals that of the output ($M=N$), ideal control means that we can shape the conductance properties of the system as we like.  In this section, we consider what happens for $M < N$.  We focus on the  example of the optimisation of the conductance of an open chaotic quantum dot \cite{Oberholzer2002} and look at both the feedback and series configurations.

\subsection{Random matrix theory}

We will use random matrix theory to describe the dot \cite{Baranger1994,Jalabert1994,Beenakker1997} and take its scattering matrix to be a $4N \times 4N$ random unitary matrix drawn from Dyson's circular ensemble.  
To study the effects of changing the size of the control space on a single system, we implement the control matrix, $K$, as a $2N \times 2N$ matrix consisting of a $2M \times 2M$ sub-matrix that represents the actual control operation, with the rest of the entries corresponding to simple reflections.
Thus, for $M=0$, the scattering matrix of the dot consists of the four-lead random $ S$ with leads C and D completely sealed off such that electrons are simply reflected back into the dot. This is the scattering matrix of the dot without control and, as such, will be used as the basis of the series calculation.
For $1 \le M \le N$, a total of $N-M$ channels reflect back into the dot and the remaining $M$ channels are scattered by the control matrix.  In this way we mimic the opening up of the dot to increase the number of channels that are affected by the controller.

We then use the controller to optimise the conductance of the dot.  Concentrating first of the feedback loop geometry, we generate the random matrix $S$, and construct the feedback matrix $S_\mathrm{fb}$ based on an arbitrary $2M\times2M$ unitary control matrix $K$ parameterised as in \citer{Zyczkowski1994}.   We then calculate the conductance of $S_\mathrm{fb}$ using \eq{EQ:cond} and numerically maximise its value over the choice of feedback controller $K$. 
This procedure is then repeated for the series geometry.

\begin{figure}[t]
  \psfrag{MN}{$M/N$}
  \psfrag{GNG0}{$G/(NG_0)$}
  \begin{center}
    \includegraphics[width=\columnwidth,clip=true]{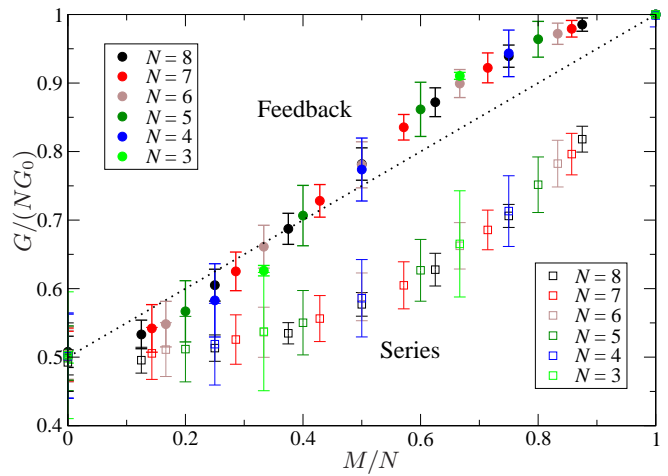}   
  \end{center}
  \caption{
    Optimised conductance $G$  of a chaotic quantum dot under coherent control in both feedback (solid circles) and series (open squares) configurations as a function of the ratio of control to output dimension, $M/N$.
    The conductance shown is the average over 100 random scattering matrices  (scaled by its maximum possible value $N G_0$) with controller $K$ chosen to maximise the conduction.  The error bars indicate the standard deviation of the conductance distribution.  
    Without control ($M=0$) the conductance takes a value of $G /(N G_0) \approx \frac{1}{2} $, in line with random matrix theory without control.
    When $M=N$, ideal control is possible for both series and feedback setups and the ballistic conductance $G/(N G_0) = 1$ is obtained.
    For $M$ increasing from $M=0$ we see a monotonic increase in the conductance for both series and feedback geometry. The feedback results, however, are clearly higher than those with series control across the entire range of $M$.
    The dotted line is a straight interpolation between start and end points: 
    $G / (NG_0)= \frac{1}{2}(1+\frac{M}{N})$.
    \label{FIG:QDFBconductance}
  }
\end{figure}

\subsection{Results}

\fig{FIG:QDFBconductance} shows the mean control-optimised conductance of 100 random $S$-matrices in both feedback and series geometries with
$2\le N \le 8$ and $0\le M \le N$.
The end points of this graph are easily understood.
For $M=0$, there is no control and no optimisation.  The average conductance is then very close to the random-matrix ensemble-average value \cite{Beenakker1997}
\beq
  G &=&  \frac{1}{2}G_0 N
  .
\eeq
The reflections used in constructing the $M=0$ scattering matrix therefore appear to give similar conductance properties to a random unitary, and this was confirmed further by examining the distribution of transmission eigenvalues for these matrices (not shown).
At the other end of the graph, for $M=N$, ideal control is possible in both feedback and series cases, and the conductance can be maximised by setting all transmission probabilities $T_{n} = 1; ~ \forall n$.  The conductance is then $G=NG_0$, which is the maximum possible for an $N$-channel conductor (the ballistic limit).  We mention that our numerical optimisation reliably finds this maximum, regardless of the starting point for the $K$-optimisation.

Between these points, the optimised conductance increases with controller size.  Interestingly, with the conductance scaled by $NG_0$ and plotted as a function of the ratio $M/N$, the optimised-conductance results for different $N$ all appear to fall on or around a single curve for each of the two geometries.
Moreover, as is clear from \fig{FIG:QDFBconductance}, in the ensemble average (and away from the known endpoints) the optimised conductance in the feedback case is always superior to that obtained from the series configuration.  In fact, in the series case, the conductance drops off rapidly as $M$ moves away from $N$, whereas the drop off for the feedback geometry is far shallower.
The maximum difference between series and feedback conductances is $\approx 0.25 N G_0$, occurring for a ratio $M/N \approx 0.63$.  Since this is fully one half the difference between the uncontrolled and ballistic conductances, the advantage of the feedback geometry in this regime is considerable.
\fig{FIG:QDFBconductance} also shows the standard deviation of the optimised-conductance distributions. In the regime, $M/N \gtrsim 0.35$, we observe that the feedback and series distributions are clearly distinct from one another.

\begin{figure}[t]
  \psfrag{GFBG0}{$G_\mathrm{fb}/G_0$}
  \psfrag{GSRG0}{$G_\mathrm{sr}/G_0$}
  \begin{center}
    \includegraphics[width=\columnwidth,clip=true]{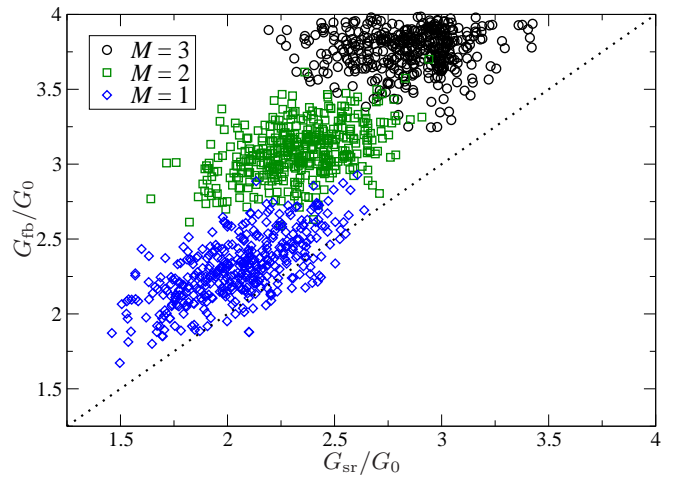}   
  \end{center}
  \caption{
    Direct comparison of the optimised conductance in series ($x$-axis) and feedback ($y$-axis) geometries for individual random matrices.
    The number of target channels was $N=4$ and results for $M=1,2,3$ are shown.
    The dotted line corresponds to $G_\mathrm{fb} = G_\mathrm{sr} $.
    For $M=1$ (blue diamonds) a few points lie below the dotted line and in these cases, the series configuration was found to be better than the feedback. In the vast majority of cases, however, the conductances with feedback exceed those of the series configuration.
    For $M=2,3$, we found $G_\mathrm{fb} > G_\mathrm{sr} $ in all cases considered, such that the feedback geometry offers a clear advantage.  This advantage increases with increasing $M < N$.
    \label{FIG:both_scatter}
  }
\end{figure}

This point is further reinforced by \fig{FIG:both_scatter} which plots the optimised conductance in the feedback configuration against that from the series configuration for individual instances of the quantum-dot scattering matrix.  

We emphasise that, for a given $M$, both series and feedback calculations have the same number of free control parameters.  
Thus, for this system at least, the feedback geometry is far more effective at conductance optimisation than the series configuration

\section{Dephasing \label{SEC:deph}}

We now study the effects of dephasing with a simple, classical dephasing model.  We will assume that transport through the plant remains phase coherent and that the controller is the only source of dephasing.  In a minimal model, we write the scattering matrix of the controller as $K \to e^{i \phi} K$ and allow the phase $\phi$ to fluctuate between $-\Delta/2$ and $\Delta/2$.  The parameter $0\le \Delta \le \pi$ is therefore a measure of the strength of the dephasing. 
With this phase in place,
the transmission block of the scattering matrix with feedback becomes $t_\mathrm{fb} \to t_\mathrm{fb}(\phi) $.  The conductance in the presence of dephasing is then calculated as
\beq
  G_\mathrm{deph.}(\Delta) 
  = 
  \frac{G_0}{\Delta}
  \int_{-\Delta /2}^{\Delta /2} d\phi
  ~
  \mathrm{Tr}
  \left[
    t^\dag_\mathrm{fb}(\phi) t_\mathrm{fb}(\phi)
  \right]
  .
\eeq
This integral can be carried out analytically by expanding the inverse in $S_\mathrm{fb}$ as geometric series, but the resulting expression can not  be resummed.  
A similar calculation can also be carried out for the series configuration.
Since the expressions so obtained are both lengthy and unenlightening, we do not reproduce them here.

\begin{figure}[t]
  \psfrag{DP}{$\Delta / \pi$}
  \psfrag{GG0}{$ G_\mathrm{deph.}/G_0$}
  \begin{center}
    \includegraphics[width=\columnwidth,clip=true]{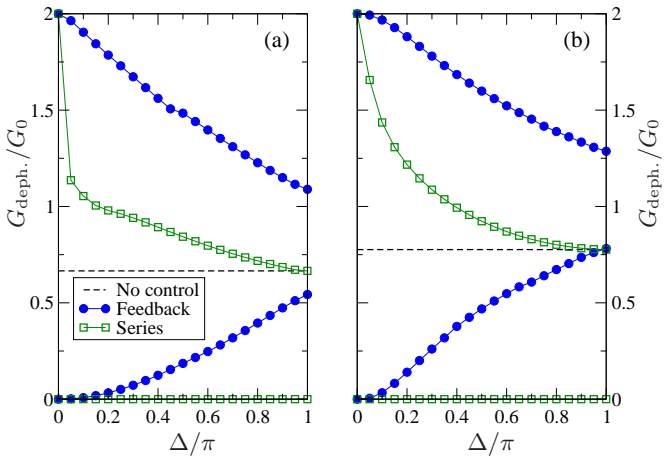}   
  \end{center}
  \caption{
    Optimised conductance $G_\mathrm{deph}$ for the chaotic dot as a function of $\Delta$ which characterises the strength of the dephasing: $\Delta =0$ corresponds to no dephasing, and $\Delta = \pi$, to a complete randomisation of the phase associated with the controller. 
    The two plots correspond to two instances of random matrix $S$.  For each we plot the minimum and maximum conductance obtained with numerical optimisation of controller $K$ in both feedback and series configurations.
    We also show the conductance with no control (dashed line).
    The most significant point is that the maximum conductance in the feedback case drops far slower as a function of $\Delta$ than does the series case.  
    Even in the completely-dephased limit,  $\Delta = \pi$, the feedback geometry offers a degree of conductance gain.
    For these plots, the channel-numbers were $N = M = 2$.
    \label{FIG:dephasing}
  }
\end{figure}

\fig{FIG:dephasing} shows results of this calculation for both series and feedback configurations with  $N = M = 2$.  We show both minimum and maximum conductance in the presence of dephasing for two particular instances of random matrix $S$ (other instances gave very similar results).
For $\Delta=0$, ideal control means that in both feedback and series cases, the maximum conductance is $G = N G_0$ and the minimum is $G=0$.  Increasing $\Delta$, the minimum conductance for the series case remains zero, since $K$ can always be set to reflect all incident electrons.
In contrast, the minimum in the feedback case increases away from zero with increasing $\Delta$.  
The maximum conductance drops as $\Delta$ increases for both cases. The drop, however, is precipitous in the series case and far more gradual in the feedback case.  Also significant is that, for $\Delta = \pi$, when the phase of the controller is completely scrambled, the maximum series conductance is reduced to its value without control (the optimum $K$ in this case is simple transmission) whereas the feedback geometry still gives a significant increase in conductance over the value without control. 
This result can be explained as follows.  Imagine an electron incident from the left in the series case. To increase transmission, paths reflected at $K$ must destructively interfere with those reflected at $S$.  This can only occur when the system is phase coherent.
On the other hand, the feedback geometry is able to increase conductance even when $S$ and $K$ are classical scatterers, since the feedback loop enables the
transmission of electrons that would otherwise have been reflected back the way they came.
In this sense, then, the series configuration is the more quantum-mechanical of the control strategies, as it relies exclusively on coherence to optimise the conduction.  Conversely, as the feedback loop has an action that can be viewed as partially classical, it is more robust in the presence of dephasing. In both cases, however, ideal control requires perfect coherence.

\section{Discussion \label{SEC:disc}}

In this paper we have introduced and studied the use of coherent control in quantum transport.
We have considered the connection of a controller to a mesoscopic scatterer in  both feedback-loop and series geometries.  In both cases we have seen that if the number of controller channels is equal to the number of output channels ($M=N$), and the controller is otherwise unconstrained, then the output scattering matrix can be set at will.

From the studies of conductance optimisation for chaotic quantum dots, two distinct advantages of the feedback geometry over the series geometry were manifest.
Firstly, away from the ideal case with $M<N$, the feedback geometry was observed to give higher conductance.  The difference between feedback and series results was substantial --- a difference of up to 50\% of the maximum possible improvement was observed.
The second advantage concerns dephasing --- the gains in conductance made by the feedback control were seen to far more robust against dephasing than in series control.
Although further investigations are necessary, we speculate that the relative advantages described here are general features of the feedback geometry and will translate to other systems.
It will also be interesting to see how these results compare with a more realistic treatment of the dephasing \cite{Seelig2001,Foerster2005}.

Our focus here has been on the optimisation of the conductance of the mesoscopic device.  The control schemes described here, however, can also be used to modify other transport properties, and in particular, the noise. 
Indeed, the suppression of current fluctuations was one of the first applications of measurement-based control in quantum transport \cite{Brandes2010}.
We have not addressed this issue directly here because, by optimising the conductance, one automatically reduces the noise.  Ideal control optimises the conductance by achieving a value of unity for all transmission probabilities, $T_n=1$.  With the zero-frequency shot noise given by \cite{Blanter2000}
$
  \mathcal{S}_\mathrm{noise}
  =
  {\textstyle \frac{2e^3 V}{h}} \sum_n T_n(1-T_n) 
$ ,
we see that optimising the conductance simply reduces the noise zero.  
Away from ideal control ($M < N$), maximisation of the conductance still results a concurrent reduction in the noise.  Direct optimisation of the noise itself would bring further gains.
More interesting will be to see how coherent control can influence the full counting statistics \cite{Levitov1996}. 
One final way in which we envisage this study could be expanded is to consider a dynamic controller, and hence the role of frequency-dependence and time-delay in coherent feedback control in quantum transport.

\acknowledgments{
JG wishes to thank EPSRC for funding under the grant EP/L006111/1Quantum Stochastic Analysis For Nanophotonic Circuit Design.
}


\end{document}